%% file: main.tex
\documentclass[letterpaper, 10 pt, journal]{IEEEtran}

\usepackage{graphicx, graphics, subfig} 
\usepackage{amsmath, amssymb, float, array, multirow, colortbl, xcolor, subcaption, mathtools}

\definecolor{candypink}{rgb}{0.89, 0.44, 0.48}
\definecolor{bittersweet}{rgb}{1.0, 0.44, 0.37}
\definecolor{bisque}{rgb}{1.0, 0.89, 0.77}
\definecolor{sunset}{rgb}{0.98, 0.84, 0.65}
\definecolor{blizzardblue}{rgb}{0.67, 0.9, 0.93}
\definecolor{babyblue}{rgb}{0.54, 0.81, 0.94}
\definecolor{lightcornflowerblue}{rgb}{0.6, 0.81, 0.93}
\definecolor{trolleygrey}{rgb}{0.5, 0.5, 0.5}
\definecolor{timberwolf}{rgb}{0.86, 0.84, 0.82}
\definecolor{lightcoral}{rgb}{0.94, 0.5, 0.5}
\definecolor{whitesmoke}{rgb}{0.96, 0.96, 0.96}
\definecolor{lavender}{rgb}{0.9, 0.9, 0.98}
\definecolor{darklavender}{rgb}{0.85, 0.85, 0.98}

\title{\LARGE \bf
Exploring Interference between Concurrent Skin Stretches}
\author{Ching Hei Cheng\textsuperscript{$\ast$}, Jonathan Eden, Denny Oetomo, Ying Tan 
\thanks{C.H. Cheng, J. Eden, D. Oetomo, Y. Tan are with the Department of Mechanical Engineering, The University of Melbourne, Australia.  \\ $\ast$ Corresponding author is C.H. Cheng. Email: chingheic@student.unimelb.edu.au}}

\begin{document}

\maketitle
\thispagestyle{empty}
\pagestyle{empty}

\begin{abstract}

Proprioception is essential for coordinating human movements and enhancing the performance of assistive robotic devices. Skin stretch feedback, which closely aligns with natural proprioception mechanisms, presents a promising method for conveying proprioceptive information. To better understand the impact of interference on skin stretch perception, we conducted a user study with 30 participants that evaluated the effect of two simultaneous skin stretches on user perception. We observed that when participants experience simultaneous skin stretch stimuli, a masking effect occurs which deteriorates perception performance in the collocated skin stretch configurations. However, the perceived workload stays the same. These findings show that interference can affect the perception of skin stretch such that multi-channel skin stretch feedback designs should avoid locating modules in close proximity.
    
\end{abstract}
\input{Introduction} %
\input{Method}

\input{Results}
\input{Discussion}

\input{Conclusion}
\section*{Acknowledgment}
This work was supported by the Australian Research Council Discovery Project (project number: DP240100938) and Valma Angliss Trust.
The authors gratefully acknowledge the contribution of Dr. Jeremy Silver (Statistical Consulting Centre, University of Melbourne) to the design, analysis, and reporting of the quantitative aspects of this research and would also like to express sincere gratitude to the participants for their valuable contributions to this research.
\bibliographystyle{ieeetr}
\bibliography{reference}
\end{document}

%% file: Introduction.tex
\section{INTRODUCTION}
Proprioception, the sense of limb position relative to the body \cite{tuthill2018proprioception}, is crucial for coordinating human movements. It plays a key role in tasks that involve multiple joints \cite{sainburg1993loss} and multiple limbs \cite{Schaffer2021Ararecase, kazennikov2005goal} working together, such as when walking and driving. However, this information is not naturally present when operating an artificial limb (e.g., a tele-operated robot or a prosthetic) and its loss has been shown to lead to reduced operator performance and increased workload \cite{schiefer2018artificial,papaleo2023integration}. The provision of supplementary proprioception back to users of artificial limbs through somatosensation, has been shown to aid the incorporation of artificial limbs into tasks and the user's body schema \cite{papaleo2023integration, blank2008identifying}. However, the capacity of users to exploit this added information is not yet completely understood.

To evaluate the benefits of providing supplementary proprioception feedback to artificial limbs users, a number of modalities such as vibrotactile \cite{noccaro2020ANovelProprioceptive, pinardi2021Cartesian, alva2020wearable}, electrotactile \cite{boljanic2022design} and skin stretch \cite{wheeler2010investigation} have been proposed. Here, vibrotactile and electrotactile feedback, which stimulate the somatosensory system through vibration and electrodermal activity, respectively, are the more compact and lightweight solutions. This has aided their integration with wearable feedback robotic systems for the artificial limbs \cite{enami2024ASurvey, pacchierotti2017wearable}. However, skin stretch, which instead stretches the skin (similar to natural proprioception mechanisms \cite{tuthill2018proprioception}) has demonstrated a lower error rates compared to other modalities \cite{bark2008comparison}. This, as well as the potential for more intuitive feedback due to the mechanism's similarity with natural proprioception, makes it a particularly attractive mechanism for legible feedback.

Proprioceptive information, like the posture of a limb, is typically multi-dimensional in nature. However, skin stretch feedback devices have been typically developed to convey only 1-dimension of information \cite{clark2018rice, bark2009wearable, schorr2013sensory}. To enable the communication of higher dimensional information through skin stretch, the typical approach has been to increase the number of skin stretch channels, where this approach has recently been applied for 3-dimensional \cite{chinello2016design, chinello2017design} and 4-dimensional feedback \cite{yem2015development}. While these devices have shown some capability to convey multi-dimensional information, the evaluation of how skin stretch understanding scales with increasing the number of channels has only been preliminary investigated \cite{zook2019effect}. Here, these initial findings have indicated that that concurrent squeezing can lead to interference with skin stretch perception.

As the skin is a continuous and deformable membrane, the activation of one skin stretch device may interfere with the perception of another, where interference could derive from the deformation of the membrane, the stimulation dynamics of the nerves or the brain's understanding of the stimuli. Interference between the same modality has been observed for vibrotactile \cite{craig1995vibrotactile} and electrotactile \cite{krishnasamy2023effect} artificial somatosensory feedback when multiple simultaneous stimuli overlap. However, interference between skin stretches has not yet been well studied, where in particular the effect of the additional stretch's placement on interference is also unclear. Here, most multi-channel skin stretch devices apply skin stretches in close proximity \cite{chinello2016design, chinello2017design, yem2015development}. However, such a configuration may lead to greater interference due to the effect of skin deformation caused by each skin stretch module. On the other hand, applying skin stretches on different limbs or on distant skin areas can eliminate the influence by skin deformation but may reduce the intuitiveness of understanding the feedback as the stimuli are not collocated \cite{congedo2006influence}. Hence, exploring different placements and quantifying the interference impact on perception and intuitiveness provide valuable data for striking a balance between perception performance and intuitiveness when designing multi-channel skin stretch feedback systems.


This study addresses the following research question: \textit{What is the impact of applying two skin stretches simultaneously on the perception and intuitiveness of skin stretch?} To investigate the interference effect and its implications for multi-channel skin stretch feedback systems, we conducted a user study that measured the perception and intuition from 30 participants who were simultaneously stimulated by two skin stretch modules (placed either in close proximity or apart) and different intensity levels. By having the two configurations, interference can be compared in scenarios where additional skin deformation in close proximity exists in one but not the other. 
Our findings provide insights into which configuration is more affected by interference in the form of just noticeable difference and what impact does the secondary stretch intensity and configuration impose on the intuitiveness of perceiving skin stretches measured by NASA-Task Load Index.

%% file: Method.tex
\section{METHOD}
\subsection{Participants}
The experiment was approved by the University of Melbourne Ethics Committee (reference number: 2024-27751-52479-5). Thirty able-bodied participants (ages 20\,-\,42, 15 female, 15 male) took part in the two-session experiment, all of whom provided informed written consent. None reported any pain or issues that could interfere with their perception of skin stretch on their forearms or affect their participation.
\subsection{Experiment Setup}
\begin{figure*}
    \centering
    \includegraphics[width=1.0\linewidth]{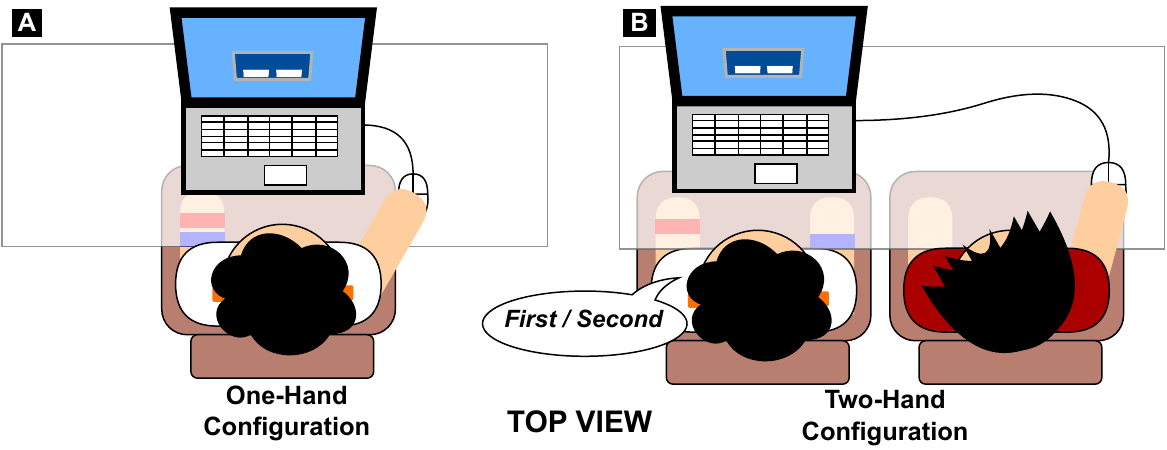}
    \caption{The experiment setup shown in the one-hand configuration \textbf{A} and the two-hand configuration \textbf{B}. The primary skin stretch module (red band) was attached to the participant's left forearm at 10\,cm from the wrist and the secondary skin stretch module (blue band) was attached 15\,cm from the wrist at the left forearm in the one-hand configuration and the right forearm in the two-hand configuration. To prevent visual and audio cues, forearms with skin stretch modules were placed underneath the opaque table and earplugs were worn. In the two-hand configuration participants voiced their responses.}
    \label{fig:experiment_setup}
\end{figure*}
\begin{figure*}
    \centering
    \includegraphics[width = 1.0\linewidth]{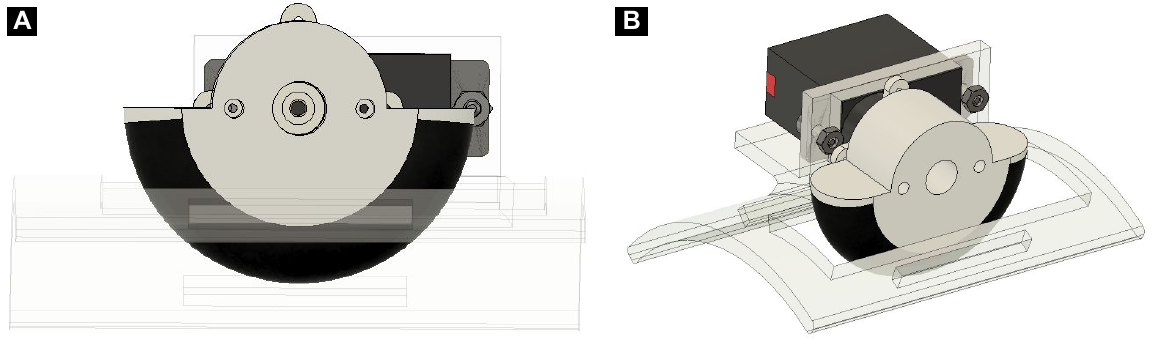}
    \caption{The skin stretch module shown in side view \textbf{A} and isometric view \textbf{B}. The thickness and diameter of the skin stretch rocker are $13\,mm$ and $40\,mm$ respectively. The black outer layer of the rocker represents the adhered silicone rubber padding.}
    \label{fig:rocker_illust}
\end{figure*}

The participants were seated comfortably in front of a computer screen throughout the experiment (Fig. \ref{fig:experiment_setup}). Two skin stretch modules (primary and secondary) were attached to their forearms: the primary module at 10 \,cm from the wrist on the dorsal side of the left forearm, and the secondary module at 15 \,cm from the wrist on the lateral side. Participants followed on-screen instructions and were asked to wear earplugs and to place their module-attached forearms beneath the table to minimize audio and visual cues during the experiment.


Two identical skin stretch modules (Fig.\,\ref{fig:rocker_illust}), designed to replicate the Rice haptic rocker \cite{battaglia2017rice, cheng2024Exploring}, were used. Each module featured a silicone rubber-padded rocker, 40 \,mm in diameter and 13 \,mm in thickness, with zero offset between the rotation axis and the rocker’s center. When the rocker rotated, it induced skin stretch in the forearm's longitudinal direction for more noticeable stretches \cite{clark2018rice}. The rockers were actuated by SC0009 servomotors (Feetech, China), with motion controlled by an Arduino Mega 2560 microcontroller and an RE-URT-01 board (Feetech, China), allowing for precise control of angular displacement and actuation time.


The experiment employed two skin stretch module configurations: \{one-hand, two-hand\}. In the one-hand configuration, the secondary skin stretch module was attached to the lateral side of the same forearm as the primary module. In the two-hand configuration, the secondary module was placed on the same position of the right forearm. Participants in this configuration were asked to voice their responses throughout the session, while those in the one-hand configuration used a computer mouse to respond.

\subsection{Protocol}
\begin{figure*}[htb!]
    \centering
    \includegraphics[width=1.0\linewidth]{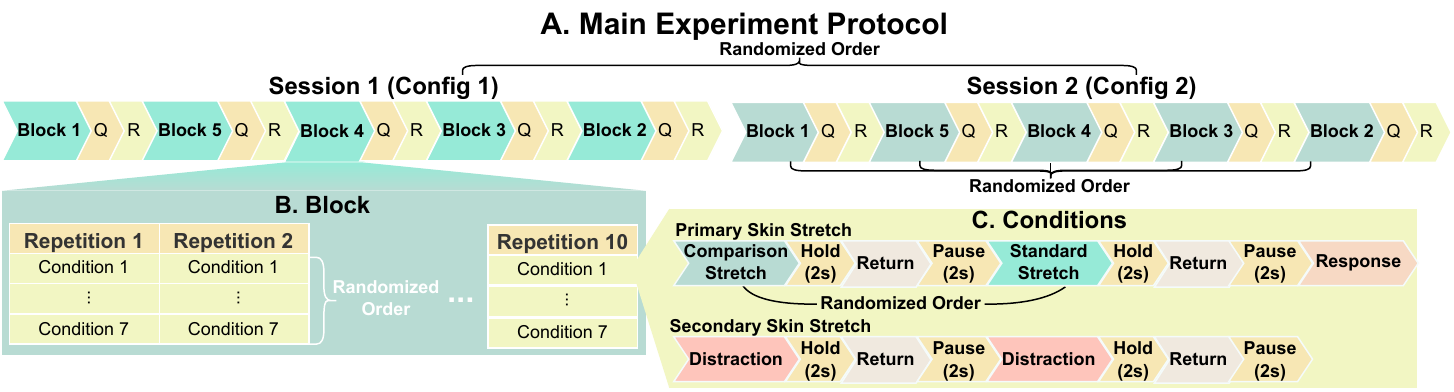}
    \caption{The experiment protocol. The main experiment protocol \textbf{A} consisted of two sessions (one-hand and two-hand), each with five blocks of varying secondary skin stretch intensity. At the end of each block, participants completed a questionnaire (Q) followed by a 5-minute rest (R). Each block (shown in \textbf{B}) contained seven different conditions, repeated ten times. In all conditions (shown in \textbf{C}), a comparison was made against a standard skin stretch. The order of blocks, conditions, and primary skin stretch was randomized.}
    \label{fig:experiment_protocol}
\end{figure*}

The experiment protocol is shown in Fig. \ref{fig:experiment_protocol}. It was conducted over two sessions, one for each configuration. Participants were divided into two groups of $15$, with one group starting in the one-hand configuration and the other in the two-hand configuration. The sessions were separated by $1$ to $7$ days (mean $2.9$ days $\pm$ standard deviation $2.03$ days).

Each session consisted of two phases: \textit{setup} and \textit{main experiment}. During the setup phase, participants were briefed about the experiment, and the skin stretch modules were attached to their forearms according to the session’s configuration. Three dummy trials were then conducted to ensure proper attachment and confirm there was no discomfort.

The main experiment phase consisted of five blocks, each testing the interference effect of a secondary skin stretch on the perception of a primary skin stretch. Each block corresponded to one of the five different secondary skin stretch levels (referred to as \textit{distraction levels}) in Table\,\ref{tab:ParaTable}. The order of the blocks was randomly assigned for each participant, ensuring that each distraction level appeared in every sequence position an equal number of times. This sequence remained constant for each participant across both experimental sessions.

%
Within each block, participants were asked to compare seven different skin stretch intensity conditions (Table\,\ref{tab:ParaTable}) delivered by the primary skin stretch module, while receiving a simultaneous secondary skin stretch. In each comparison, the primary skin stretch module delivered two consecutive stretches: one as the \textit{standard stretch} and the other as the \textit{comparison stretch}. The secondary skin stretch module instead provided the same distraction stretch concurrently with each primary stretch of the comparison. After each comparison, participants indicated which stretch from the primary skin stretch module they perceived as stronger. Each skin stretch had a constant actuation time of 0.73\,s, and all conditions were compared to the default stretch ten times. The order of conditions within each repetition was randomized.

\begin{table}[htbp]
    \centering
    \begin{tabular}{|c|ccccccc|}
    \hline
    \rowcolor{timberwolf} \textbf{Conditions} & \textbf{1} & \textbf{2} & \textbf{3} & \textbf{4*} & \textbf{5} & \textbf{6} & \textbf{7} \\ 
    \hline
    \rule{0pt}{10pt}\textbf{Primary angular} & \multirow{2}{*}{24} & \multirow{2}{*}{26} & \multirow{2}{*}{28} & \multirow{2}{*}{30} & \multirow{2}{*}{32} & \multirow{2}{*}{34} & \multirow{2}{*}{36} \\\textbf{displacement (\(^o\))} & & & & & & & \\[2.5pt]
    \hline
    \rowcolor{darklavender} \textbf{Distraction levels} & &\textbf{1} & \textbf{2} & \textbf{3} & \textbf{4} & \textbf{5} & \\
    \hline
    \rule{0pt}{10pt}\textbf{Secondary angular} & & \multirow{2}{*}{-12} & \multirow{2}{*}{-6} & \multirow{2}{*}{0} & \multirow{2}{*}{6} & \multirow{2}{*}{12} & \\
    \textbf{displacement (\(^o\))} & & & & & & & \\[2.5pt]
    \hline
    \end{tabular}

    \caption{Angular displacement parameters of the seven skin stretch conditions delivered by the primary skin stretch module, where * denotes the \textit{standard stretch}, and the five secondary skin stretch module distraction levels.}
    \label{tab:ParaTable}
\end{table}
%
After each block, participants completed a questionnaire (\textbf{Q} in Fig.\,\ref{fig:experiment_protocol}) to assess the workload associated with that block. Participants were then given a 5-minute break between blocks.

\subsection{Data Analysis}
Participant performance was evaluated based on two criteria: their perception, measured by the just noticeable difference (JND), and their intuitiveness, using the NASA-TLX \cite{hart2006nasa}.
\subsubsection{Just Noticeable Difference (JND)}
\begin{figure}[h]
    \centering
    \includegraphics[width=1.0\linewidth]{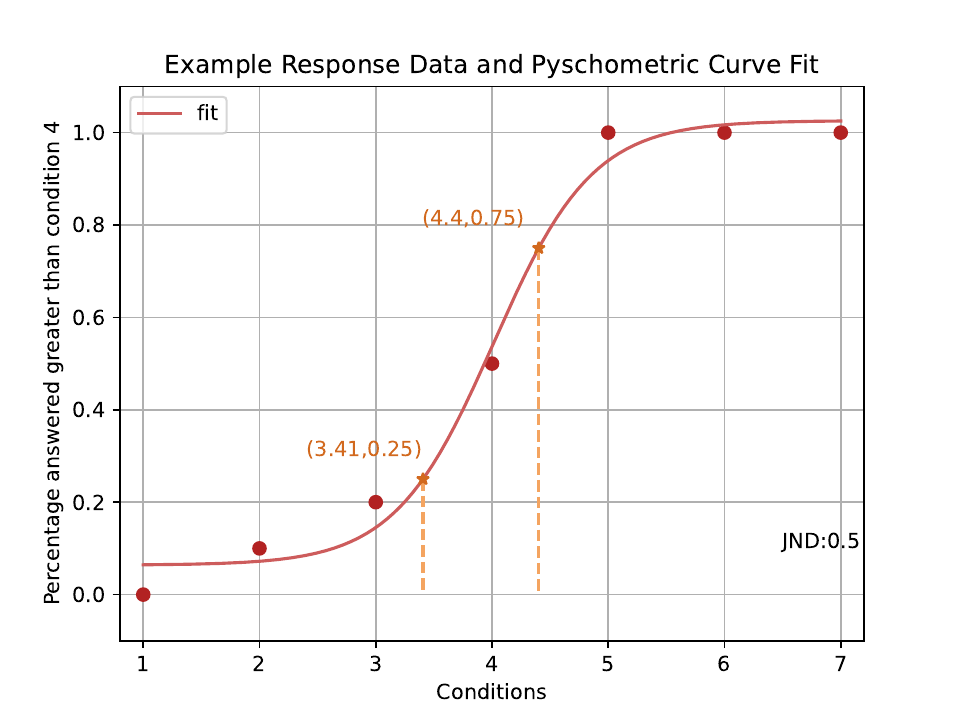}
    \caption{Example response data and fitted psychometric curve for JND calculation. The $25\%$ \((DL_{l})\) and $75\%$ \((DL_{u})\) points (denoted by stars) were used to compute the JND as in (\ref{eq: JND}).}
    \label{fig:example_psy_curve_fit}
\end{figure}

The just noticeable difference (JND), defined as the smallest change in a stimulus needed for it to be perceived as different from another stimulus \cite{levine1991fundamentals}, was measured using the method of constant stimuli. In this method, one stimulus in each pair was fixed as the standard stretch, as described in \cite{gescheider2013psychophysics}.
%
%
As shown in Fig.\,\ref{fig:example_psy_curve_fit}, the JND was then computed by plotting the proportion of greater responses against the comparison stimuli values. Here, a logistic curve was fit to this data to obtain the psychometric curve for the skin stretch with the given standard stimulus. The intermediate proportion points corresponding to stimuli with $25\%$ and $75\%$ greater responses were then used to compute the JND, as indicated by \cite{gescheider2013psychophysics},
\begin{equation}
    JND = \dfrac{(DL_{u} - DL_{l})}{2},
    \label{eq: JND}
\end{equation}
where \(DL_{u}\) and \(DL_{l}\) are the values where the psychometric curve has 75\% and 25\% of greater responses.

\subsubsection{NASA-Task Load Index (NASA-TLX)}

The perceived task load was measured using the NASA-TLX questionnaire, an assessment tool for perceived workload \cite{hart1988development}. This was applied to evaluate the intuitiveness of differentiating skin stretches under the two configurations and five secondary skin stretch conditions. NASA-TLX includes six subcategories: i) mental demand; ii) physical demand; iii) temporal demand; iv) performance; v) effort; and vi) frustration. A single raw NASA-TLX score was calculated by summing all the subcategory scores with equal weighting and remapping to a scale of $100$.

\subsubsection{Statistical Analysis}
Mixed-effects models \cite{raudenbush2002hierarchical} were used to analyze the JND and NASA-TLX data. The configuration of the skin stretch modules and the distraction level of the secondary skin stretch were treated as fixed effects, while the participant identifier was treated as a random effect. A significance threshold of $0.05$ was set for all statistical tests.


To apply mixed-effects modeling, we first checked the data distribution for both the NASA-TLX and JND, as they must meet the homoscedasticity assumption, which requires that the variance within each group of data is roughly equal. The NASA-TLX data followed a normal distribution and satisfied the assumption. However, the JND data did not meet this requirement. To address this, the JND data was normalised and analyzed using a Beta distribution, which provided a better fit. Consequently, an ordered beta regression mixed-effects model was used for the JND data, while a Gaussian regression mixed-effects model was applied to the NASA-TLX data. Then, a repeated measures Two-way ANOVA (for NASA-TLX) or Wald's test (for JND) was used for finding out whether the fixed effects have significance impact on the metrics or not. When fixed effects were found to have a clear effect on the metric, a pairwise comparison (Z-test for the ordered beta regression mixed effect model and T-test for the Gaussian regression mixed effect model) with the Tukey adjustment for multiple comparisons was used.


%% file: Results.tex
\section{RESULTS}
This section discusses participant performance using JND, as well as their intuitiveness, as measured by the NASA-TLX.
\subsection{Performance}
\begin{figure*}[h!]
    \centering
    \includegraphics[width = \linewidth]{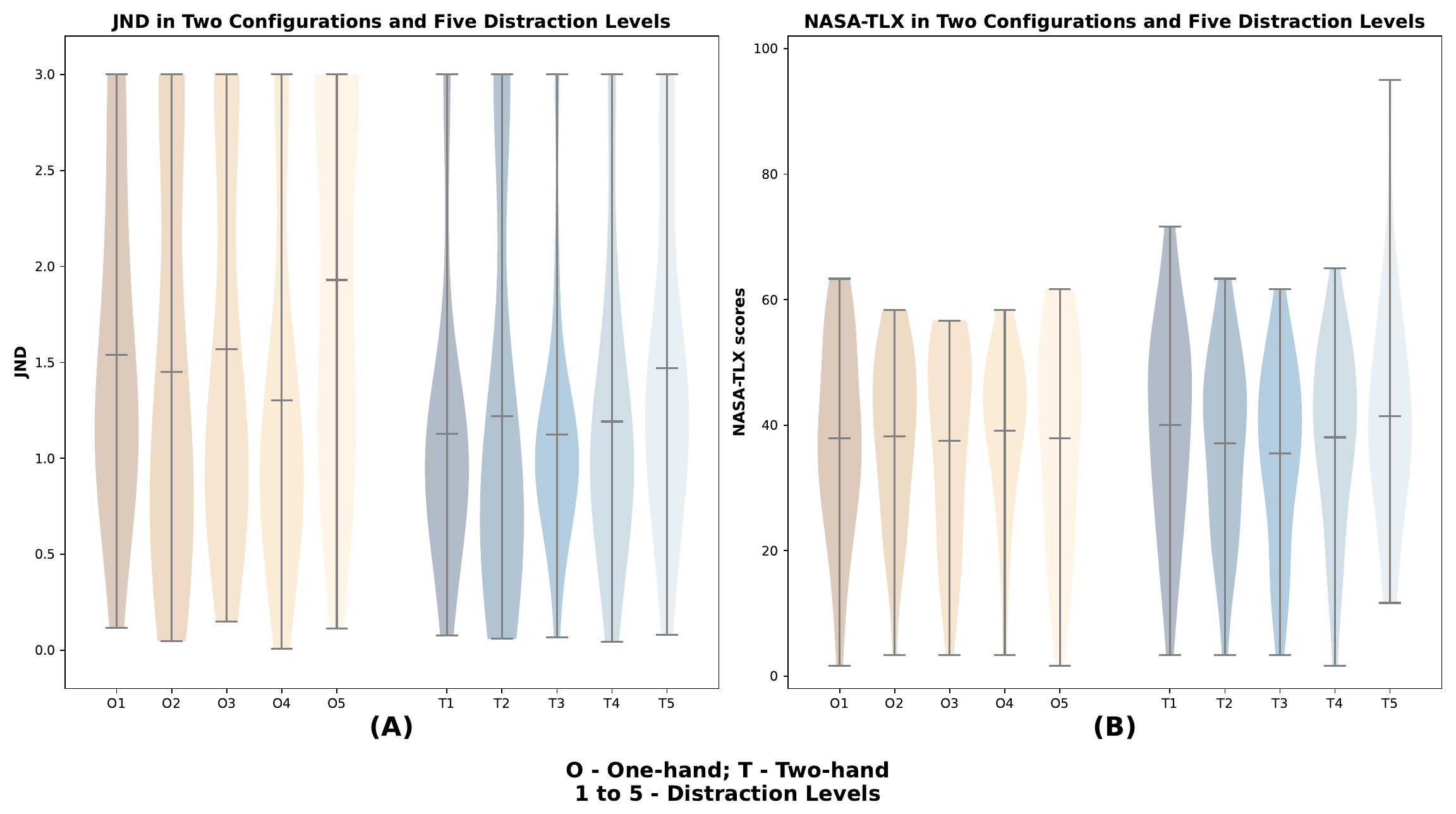}
    \caption{JND and NASA-TLX results plotted by configuration and distraction levels. The results are shown as a violin plot, where the shape represents the data distribution and the middle bars represents the mean. The yellow and the blue colors correspond to the one-hand and two-hand configuration and the shades depict the distraction levels within the same configuration.}
    \label{fig:combine_violin_plot}
\end{figure*}

Fig.\,\ref{fig:combine_violin_plot}A depicts the JND results for each condition and distraction level. The JND was found to be affected by both the configuration (Wald's $\chi^2$ test: \(\chi(1)^2 = 7.610,\,p = 0.0058\)) and the distraction level (\(\chi(4)^2 = 11.716,\,p = 0.0196\)) but not by the interaction of the two factors (\(\chi(4)^2 = 0.944,\,p = 0.9182\). Here, while the JND was higher in the one-hand configuration then in the two-hand configuration, post-hoc analysis of the effect of the distraction level indicated that the only pair that showed a clear difference was between distraction level two and five (Z-test: \(z = -3.255,\,p = 0.01\)).

These results indicate that participants perceived finer changes when skin stretches were stimulating on different limbs (mean JND\(\,=\,2.4530^\circ\)) compared to collocated skin areas (mean JND\(\,=3.1168^\circ\)). Moreover, there did not appear to be a clear effect of the distraction level with the exception of differences with distraction level 5 (corresponding to a positive displacement of the secondary skin stretch device by $12^\circ$). 

\subsection{Intuitiveness}
The NASA-TLX is shown for the different configuration and distraction levels in Fig.\,\ref{fig:combine_violin_plot}B.
No clear effect was observed for the configuration (Repeated measures two-factor ANOVA: \(F(1,292) = 0.073,\,p = 0.7870\)), distraction levels (\(F(4,292) = 1.151,\,p = 0.3327\)) or their interaction (\(F(4,288) = 1.090,\,p = 0.3615\)).

This suggests that the workload for humans to perceive skin stretches in either collocated or on different limbs were not clearly different nor was there a clear difference caused by the the intensity of the skin stretches.

%% file: Discussion.tex
\section{DISCUSSION}

This study investigated the impact of adding a second skin stretch stimulus on the perception of a primary skin stretch. The secondary skin stretch was tested in two attachment configurations (one-hand and two-hand) and at five different distraction intensities. Just Noticeable Difference (JND) and NASA-TLX scores were measured. The results showed that the two-hand configuration had a lower JND compared to the one-hand configuration, while no clear difference in NASA-TLX scores was observed between the two configurations. Additionally, the intensity factor did not show a clear trend in its effect on JND and did not impact the NASA-TLX scores.

The difference in perception performance suggests that there was some interference in the user's perception due to the secondary stretch. This interference could arise from three possible sources: (i) the brain having difficulty to integrate the different feedback signals, (ii) the nervous system introducing interference, and (iii) the skin deformation changing the mechanics of the feedback. At the brain level, one possibility is that the human brain may have been unable to process both skin stretch signals simultaneously. However, since the mean JND for the two-hand configuration was approximately the same as the single skin stretch setup in \cite{cheng2024Exploring}, this indicates that the human brain was capable of perceiving both skin stretches without affecting the JND of the primary stretch. At the nerve level, distraction level 3, which had no secondary stretch, showed no difference compared to other distraction levels, making it unlikely to be the primary cause. In addition, the one-hand level 3 had a noticeably higher mean JND than that from both the two-hand configuration level 3 condition and that observed previously with only one skin-stretch device \cite{cheng2024Exploring}, where these two cases did not have a clear difference in JND. Therefore, the most likely contributing factor appears to be at the skin level, where the collocated skin stretch modules may have resulted in a complex skin deformation, causing the receptors to receive interfering signals. However, further studies would be needed to support this hypothesis.

One interesting observation from the results was that, although participants performed worse in terms of JND in the one-hand configuration, their perceived workload (as measured by NASA-TLX) remained unchanged. This suggests that participants perceived both configurations as equally difficult, potentially indicating that the configuration difference did not impact the intuitiveness of the feedback. Other than NASA-TLX, response time or task completion time are metrics that were also used in studies to reflect the intuitiveness of the feedback. In this work, although the  total session time was recorded and no clear difference was observed across different configuration and intensities, the response mechanisms of one-hand and two-hand configuration are different and hence, such metric is imperfect for fair comparison. For future studies, measuring participants' response times (rather then the cumulative session time) with equal response mechanisms could provide clearer insight into how the intuitiveness of the skin stretch feedback may evolve.

For the distraction levels, only levels 2 and 5 showed a clear difference where level 5 had a higher mean JND. The reason for this is unclear, as these levels differ in both the direction and intensity of the secondary stretch, and no clear trend emerged. The distraction intensities might have been too close to the sensation threshold, preventing a noticeable effect. However, from Fig.\ref{fig:combine_violin_plot}, both level 5 seems to have a slightly higher mean than other intensities in the same configuration, hinting that there might be a potential trend here. Increasing the intensity or matching it to the primary skin stretch might make the interference effect more apparent, as \cite{zook2019effect} noted that stronger forces can mask weaker ones. Thus, the interference could be more significant when the two skin stretches are of similar force or torque levels.

Our findings suggest that for skin stretch-based artificial feedback devices, providing the devices in close proximity on the same limb may negatively affect perception. While most of the current multi-channel skin stretch devices apply skin stretch stimuli in this configuration for more intuitive understanding of the information, there was no clear difference in the NASA-TLX result with the devices on separate limbs, suggesting that this may be a viable solution for higher dimensional skin stretch. However, the limb-separated stimuli approach is limited by the number of limbs available. An alternative to avoid interference may instead be to place the skin stretch cues further apart on the same limb or to consider using other modalities (such as electrotactile or vibrotactile feedback) to extend the device's capabilities. For the former option, further studies are needed to determine the optimal distance between additional channels and to explore the effects of following multiple signals simultaneously. For the latter, care should be taken to assess how different modalities interact and affect user performance.



%% file: Conclusion.tex
\section{CONCLUSION}
This study investigated the interference effect between two skin stretch channels, testing two attachment configurations and five distraction stretch intensities using rocker-based skin stretch modules. The two-hand configuration resulted in a lower JND for the primary skin stretch, but neither configuration nor distraction intensity affected participant workload. Future work could explore higher distraction intensities to assess broader interaction effects, and introduce response time metrics with an equivalent response mechanism in both configurations to support the measurement of intuitiveness.

%% file: main.bbl
\begin{thebibliography}{10}

\bibitem{tuthill2018proprioception}
J.~C. Tuthill and E.~Azim, ``Proprioception,'' {\em Curr. Biol.}, vol.~28, no.~5, pp.~R194--R203, 2018.

\bibitem{sainburg1993loss}
R.~L. Sainburg, H.~Poizner, and C.~Ghez, ``Loss of proprioception produces deficits in interjoint coordination,'' {\em J. Neurophysiol.}, vol.~70, no.~5, pp.~2136--2147, 1993.

\bibitem{Schaffer2021Ararecase}
J.~E. Schaffer, F.~R. Sarlegna, and R.~L. Sainburg, ``A rare case of deafferentation reveals an essential role of proprioception in bilateral coordination,'' {\em Neuropsychologia}, vol.~160, p.~107969, 2021.

\bibitem{kazennikov2005goal}
O.~V. Kazennikov and M.~Wiesendanger, ``Goal synchronization of bimanual skills depends on proprioception,'' {\em Neurosci. Lett.}, vol.~388, no.~3, pp.~153--156, 2005.

\bibitem{schiefer2018artificial}
M.~A. Schiefer, E.~L. Graczyk, S.~M. Sidik, D.~W. Tan, and D.~J. Tyler, ``Artificial tactile and proprioceptive feedback improves performance and confidence on object identification tasks,'' {\em PLOS ONE}, vol.~13, pp.~1--18, 12 2018.

\bibitem{papaleo2023integration}
E.~D. Papaleo, M.~D’Alonzo, F.~Fiori, V.~Piombino, E.~Falato, F.~Pilato, A.~De~Liso, V.~Di~Lazzaro, and G.~Di~Pino, ``Integration of proprioception in upper limb prostheses through non-invasive strategies: a review,'' {\em J. Neuroeng. Rehabilitation}, vol.~20, no.~1, p.~118, 2023.

\bibitem{blank2008identifying}
A.~Blank, A.~M. Okamura, and K.~J. Kuchenbecker, ``Identifying the role of proprioception in upper-limb prosthesis control: Studies on targeted motion,'' {\em ACM Trans. Appl. Percept.}, vol.~7, no.~3, pp.~1--23, 2008.

\bibitem{noccaro2020ANovelProprioceptive}
A.~Noccaro, L.~Raiano, M.~Pinardi, D.~Formica, and G.~D. Pino, ``A novel proprioceptive feedback system for supernumerary robotic limb,'' in {\em IEEE RAS EMBS Int. Conf. Biomed. Robot. Biomechatron.}, pp.~1024--1029, 2020.

\bibitem{pinardi2021Cartesian}
M.~Pinardi, L.~Raiano, A.~Noccaro, D.~Formica, and G.~Di~Pino, ``Cartesian space feedback for real time tracking of a supernumerary robotic limb: a pilot study,'' in {\em Int. IEEE/EMBS Conf. Neural Eng. NER}, pp.~889--892, 2021.

\bibitem{alva2020wearable}
P.~G.~S. Alva, S.~Muceli, S.~F. Atashzar, L.~William, and D.~Farina, ``Wearable multichannel haptic device for encoding proprioception in the upper limb,'' {\em J. Neural Eng.}, vol.~17, p.~056035, Oct 2020.

\bibitem{boljanic2022design}
T.~Boljani{\'c}, M.~Isakovi{\'c}, J.~Male{\v{s}}evi{\'c}, D.~Formica, G.~Di~Pino, T.~Keller, and M.~{\v{S}}trbac, ``Design of multi-pad electrotactile system envisioned as a feedback channel for supernumerary robotic limbs,'' {\em Artif. Organs}, vol.~46, no.~10, pp.~2034--2043, 2022.

\bibitem{wheeler2010investigation}
J.~Wheeler, K.~Bark, J.~Savall, and M.~Cutkosky, ``Investigation of rotational skin stretch for proprioceptive feedback with application to myoelectric systems,'' {\em IEEE Trans. Neural Syst. Rehabilitation Eng.}, vol.~18, no.~1, pp.~58--66, 2010.

\bibitem{enami2024ASurvey}
M.~Emami, A.~Bayat, R.~Tafazolli, and A.~Quddus, ``A survey on haptics: Communication, sensing and feedback,'' {\em IEEE Commun. Surv. Tutor.}, pp.~1--1, 2024.

\bibitem{pacchierotti2017wearable}
C.~Pacchierotti, S.~Sinclair, M.~Solazzi, A.~Frisoli, V.~Hayward, and D.~Prattichizzo, ``Wearable haptic systems for the fingertip and the hand: taxonomy, review, and perspectives,'' {\em IEEE Trans. Haptics}, vol.~10, no.~4, pp.~580--600, 2017.

\bibitem{bark2008comparison}
K.~Bark, J.~W. Wheeler, S.~Premakumar, and M.~R. Cutkosky, ``Comparison of skin stretch and vibrotactile stimulation for feedback of proprioceptive information,'' in {\em IEEE Symp. Haptic Interfaces Virtual Environ. Teleoperator Syst.}, pp.~71--78, 2008.

\bibitem{clark2018rice}
J.~P. Clark, S.~Y. Kim, and M.~K. O’Malley, ``The {RICE} haptic rocker: Comparing longitudinal and lateral upper-limb skin stretch perception,'' in {\em EuroHaptics: Int. Conf. Hum. Haptic Sens. Touch Enabled Comput. Appl.}, pp.~125--134, 2018.

\bibitem{bark2009wearable}
K.~Bark, J.~Wheeler, G.~Lee, J.~Savall, and M.~Cutkosky, ``A wearable skin stretch device for haptic feedback,'' in {\em EuroHaptics Conf. Symp. Haptic Interfaces Virtual Environ. Teleoperator Syst.}, pp.~464--469, IEEE, 2009.

\bibitem{schorr2013sensory}
S.~B. Schorr, Z.~F. Quek, R.~Y. Romano, I.~Nisky, W.~R. Provancher, and A.~M. Okamura, ``Sensory substitution via cutaneous skin stretch feedback,'' in {\em IEEE Int. Conf. Robot. Autom.}, pp.~2341--2346, 2013.

\bibitem{chinello2016design}
F.~Chinello, C.~Pacchierotti, N.~G. Tsagarakis, and D.~Prattichizzo, ``Design of a wearable skin stretch cutaneous device for the upper limb,'' in {\em IEEE Haptics Symp.}, pp.~14--20, 2016.

\bibitem{chinello2017design}
F.~Chinello, C.~Pacchierotti, J.~Bimbo, N.~G. Tsagarakis, and D.~Prattichizzo, ``Design and evaluation of a wearable skin stretch device for haptic guidance,'' {\em IEEE Robot. Autom. Lett.}, vol.~3, no.~1, pp.~524--531, 2017.

\bibitem{yem2015development}
V.~Yem, M.~Otsuki, and H.~Kuzuoka, ``Development of wearable outer-covering haptic display using ball effector for hand motion guidance,'' {\em Haptic Interaction: Perception, Devices and Applications}, pp.~85--89, 2015.

\bibitem{zook2019effect}
Z.~A. Zook, J.~J. Fleck, T.~W. Tjandra, and M.~K. O’Malley, ``Effect of interference on multi-sensory haptic perception of stretch and squeeze,'' in {\em IEEE World Haptics Conf}, pp.~371--376, 2019.

\bibitem{craig1995vibrotactile}
J.~C. Craig, ``Vibrotactile masking: The role of response competition,'' {\em Percept. Psychophys.}, vol.~57, no.~8, pp.~1190--1200, 1995.

\bibitem{krishnasamy2023effect}
J.~Krishnasamy~Balasubramanian, R.~K. Ray, and M.~Muniyandi, ``Effect of subthreshold electrotactile stimulation on the perception of electrovibration,'' {\em ACM Trans. Appl. Percept.}, vol.~20, no.~3, pp.~1--16, 2023.

\bibitem{congedo2006influence}
M.~Congedo, A.~L{\'e}cuyer, and E.~Gentaz, ``The influence of spatial delocation on perceptual integration of vision and touch,'' {\em Presence}, vol.~15, no.~3, pp.~353--357, 2006.

\bibitem{battaglia2017rice}
E.~Battaglia, J.~P. Clark, M.~Bianchi, M.~G. Catalano, A.~Bicchi, and M.~K. O'Malley, ``The {R}ice haptic rocker: skin stretch haptic feedback with the {P}isa/{IIT} softhand,'' in {\em IEEE World Haptics Conf}, pp.~7--12, 2017.

\bibitem{cheng2024Exploring}
C.~H. Cheng, J.~Eden, D.~Oetomo, and Y.~Tan, ``Exploring the influence of displacement, velocity and actuation duration on skin stretch perception,'' in {\em IEEE RAS EMBS Int. Conf. Biomed. Robot. Biomechatron.}, pp.~229--234, 2024.

\bibitem{hart2006nasa}
S.~G. Hart, ``{NASA}-task load index ({NASA-TLX}); 20 years later,'' in {\em Proc Hum Factors Ergon Soc Annu Meet}, vol.~50, pp.~904--908, Sage publications Sage CA: Los Angeles, CA, 2006.

\bibitem{levine1991fundamentals}
M.~W. Levine and J.~M. Shefner, {\em Fundamentals of Sensation and Perception}.
\newblock Brooks/Cole Publishing Company, 1991.

\bibitem{gescheider2013psychophysics}
G.~A. Gescheider, {\em Psychophysics: the fundamentals}.
\newblock Psychology Press, 2013.

\bibitem{hart1988development}
S.~G. Hart and L.~E. Staveland, ``Development of {NASA-TLX} (task load index): Results of empirical and theoretical research,'' in {\em Advances in Psychology}, vol.~52, pp.~139--183, Elsevier, 1988.

\bibitem{raudenbush2002hierarchical}
S.~W. Raudenbush, ``Hierarchical linear models: Applications and data analysis methods,'' {\em Advanced Quantitative Techniques in the Social Sciences Series}, 2002.

\end{thebibliography}
